\documentclass[conference,a4paper]{IEEEtran}

\newtheorem{thm}{Theorem}

\newtheorem{lem}{Lemma}

\newtheorem{defn}{Definition}

\usepackage[final]{graphicx}
\usepackage[reqno]{amsmath}
\usepackage{amssymb}
\usepackage{algorithm}
\usepackage{algorithmicx}
\usepackage{algpseudocode}
\usepackage{setspace}
\usepackage{multicol}
\usepackage{comment}

\begin{document}

\sloppy

\title{The Role of Transmitter Cooperation in Linear Interference Networks with Block Erasures}
\author{\IEEEauthorblockN{Yasemin Karacora, Tolunay Seyfi and Aly El Gamal}
 \IEEEauthorblockA{ECE Department, Purdue University\\ Email: \{ykaracor,tseyfi,elgamala\}@purdue.edu}}

\maketitle

\begin{abstract}
In this work, we explore the potential and optimal use of transmitter cooperation in wireless interference networks with deep fading conditions. We consider a linear interference network with $K$ transmitter-receiver pairs, where each transmitter can be connected to two neighboring receivers. Long-term fluctuations (shadow fading) in the wireless channel can lead to any link being erased with probability $p$. Each receiver is interested in one unique message that can be available at two transmitters. The considered rate criterion is the average per user degrees of freedom (puDoF) as $K$ goes to infinity. Prior to this work, the optimal assignment of messages to transmitters were identified in the two limits $p \rightarrow 0$ and $p \rightarrow 1$. We identify new schemes that achieve average puDoF values that are higher than the state of the art for a significant part of the range $0 < p < 1$. The key idea to our results is to understand that the role of cooperation shifts from increasing the probability of delivering a message to its intended destination at high values of $p$, to interference cancellation at low values of $p$. Our schemes are based on an algorithm that achieves the optimal DoF value in any network realization, when restricted to a given message assignment as well as the use of zero-forcing schemes.
\end{abstract}

\section{Introduction}
Our focus in this work is to analyze information theoretic models of interference networks that capture the effect of deep fading conditions through introducing random link erasure events in blocks of communication time slots. More specifically, in order to consider the effect of long-term fluctuations (deep fading or shadowing), we assume that communication takes place over blocks of time slots, and independent link erasures take place with a probability $p$ in each block. Further, short-term channel fluctuations allow us to assume that in each time slot, all non erased channel coefficients are drawn independently from a continuous distribution; this is known as the assumption that the channel is \emph{generic}. 

We are interested in understanding the role of transmitter cooperation (also known as Coordinated Multi-Point (CoMP) Transmission) in these dynamic interference networks. In particular, if each message can be assigned to more than one transmitter, with a restriction only on the maximum number of such transmitters, without any constraint on their identity, what would be the optimal assignment of messages to transmitters and corresponding transmission scheme that maximizes the average rate over all possible realizations of the network? To simplify analysis, we consider the linear interference network introduced in~\cite{Wyner}, where each transmitter can only be connected to the receiver having the same index as well as one following receiver. The channel capacity criterion we consider is the pre-log factor of the sum capacity at high Signal to Noise Ratio (SNR), also known as the Degrees of Freedom (DoF). Because our goal is to understand the optimal pattern of transmitter cooperation that scales in large networks, we consider the DoF normalized by the number of transmitter-receiver pairs, and take the limit as that number goes to infinity; we call this the per user degrees of freedom (puDoF). 

In~\cite{ElGamal-Veeravalli-Asilomar13}, the considered setting was studied, where first the case where each message can only be available at a single transmitter was analyzed. The optimal assignment of messages to transmitters, and the value of the average puDoF were identified as a function of the erasure probability $p$. 
In this work, we extend the work of~\cite{ElGamal-Veeravalli-Asilomar13} by studying the case where each message can be available at two transmitters, and transmitter cooperation is allowed. The optimal message assignment in the limits $p \rightarrow 1$ and $p \rightarrow 0$ were identified in~\cite{ElGamal-Veeravalli-Asilomar13} and~\cite{ElGamal-Annapureddy-Veeravalli-ICC12}, respectively. As $p \rightarrow 1$, each message is assigned to the two transmitters connected to its destination, to maximize the probability of successful delivery. As $p \rightarrow 0$, the puDoF value goes to $\frac{4}{5}$, and is achieved by splitting the network into subnetworks; each has five transmitter-receiver pairs. In order to avoid interference between the subnetworks, the last transmitter in each subnetwork is inactive. And hence, each of the first and last messages in each subnetwork is only assigned to one of the two transmitters connected to its destination, and the other assignment is used at a transmitter not connected to its destination, but connected to another receiver that is prone to interference caused by this message. Further, the middle message in each subnetwork is not transmitted. We find, through simulations, in this work that assigning that middle message to only one transmitter connected to its destination, and another transmitter not connected to its destination, leads to better rates than assigning it to the two transmitters connected to its destination at low values of $p$. That implies that a fraction of $\frac{3}{5}$ of the messages are assigned to only one of the two transmitters connected to their destination, and the remaining $\frac{2}{5}$ are assigned to the two transmitters connected to their destination. We show in this work, that at any value of $p$ from $0$ to $1$, the assignment achieving the highest puDoF using an optimal zero-forcing scheme, has a fraction of $f(p)$ of messages that are assigned to only one of the transmitters connected to their destination, and another transmitter used for interference cancellation, and the remaining fraction $1-f(p)$ of messages are assigned to the two transmitters connected to their destination. The value of $f(p)$ decreases monotonically from $\frac{3}{5}$ to $0$ as $p$ increases from $0$ to $1$, which agrees with the intuition about the shifting role of cooperative transmission from canceling interference to increasing the probability of successful delivery as $p$ increases from $0$ to $1$.  

\section{System Model and Notation}\label{sec:systemmodel}
We use the standard model for the $K-$user interference channel with single-antenna transmitters and receivers,
\begin{equation}
Y_i(t) = \sum_{j=1}^{K} H_{i,j}(t) X_j(t) + Z_i(t),
\end{equation}
where $t$ is the time index, $X_j(t)$ is the transmitted signal of transmitter $j$, $Y_i(t)$ is the received signal at receiver $i$, $Z_i(t)$ is the zero mean unit variance Gaussian noise at receiver $i$, and $H_{i,j}(t)$ is the channel coefficient from transmitter $j$ to receiver $i$ over the time slot $t$. We remove the time index in the rest of the paper for brevity unless it is needed. Finally, we use $[K]$ to denote the set $\{1,2,\ldots,K\}$

\subsection{Channel Model}
Each transmitter can only be connected to its corresponding receiver as well as one following receiver, and the last transmitter can only be connected to its corresponding receiver.
In order to consider the effect of long-term fluctuations (shadowing), we assume that communication takes place over blocks of time slots, and let $p$ be the probability of block erasure. In each block, we assume that for each $j$, and each $i \in \{j,j+1\}$, $H_{i,j}=0$ with probability $p$. Moreover, short-term channel fluctuations allow us to assume that in each time slot, all non-zero channel coefficients are drawn independently from a continuous distribution. Finally, we assume that global channel state information is available at all nodes. 
\subsection{Message Assignment}
For each $i \in [K]$, let $W_i$ be the message intended for receiver $i$, and ${\cal T}_i \subseteq [K]$ be the transmit set of receiver $i$, i.e., those transmitters with the knowledge of $W_i$. The transmitters in ${\cal T}_i$ cooperatively transmit the message $W_i$ to the receiver $i$. The messages $\{W_i\}$ are assumed to be independent of each other. Each message can only be available at two transmitters,
\begin{equation}
|{\cal T}_i| \leq 2, \forall i\in[K].
\end{equation}

\subsection{Degrees of Freedom}
The total power constraint across all the users is $P$. In each block of time slots, the rates $R_i(P)$ are achievable if the decoding error probabilities of all messages can be simultaneously made arbitrarily small as the block length goes to infinity, and this holds for almost all realizations of non-zero channel coefficients. The sum capacity $\mathcal{C}_{\Sigma}(P)$ is the maximum value of the sum of the achievable rates. The total number of degrees of freedom ($\eta$) is defined as $\limsup_{P \rightarrow \infty}\frac{ C_{\Sigma}(P)}{\log P}$. For a $K$-user channel, and a probability of block erasure $p$, we let $\eta_p(K)$ be the average value of $\eta$ over possible choices of non-zero channel coefficients.
We further define the asymptotic per user DoF (puDoF) $\tau_p$ to measure how $\eta_p(K)$ scales with $K$. 
\begin{equation}
\tau_p = \lim_{K\rightarrow \infty} \frac{\eta_p(K)}{K}
\end{equation}

\subsection{Zero-forcing (Interference Avoidance) Schemes}
We consider in this work the class of \textit{interference avoidance} schemes, where all interference is cancelled \emph{over the air}. Each message is either not transmitted or allocated one degree of freedom.
Accordingly, every receiver is either active or inactive. An active receiver does not observe interfering signals.

\section{Optimal Zero-Forcing Scheme}
\label{sec:zero}
We make the following definition of a cluster of users within the $K$-user network.
\begin{defn}
We say that a set of users with consecutive $N$ indices (say having indices in the set $[N]=\{1,2,\cdots,N\}$) form a cluster if all the diagonal links exist, i.e., $H_{i+1,i} \neq 0, \forall i\in[N-1]$, and the diagonal link between the last transmitter in the cluster and the following receiver is erased, i.e., $H_{N+1,N}=0$.
\end{defn}

A cluster as defined above is given as an input to Algorithm 1. The output of the algorithm is the transmit signals $\{X_i,i\in[N]\}$ that employs zero-forcing transmit beamforming to maximize the DoF value for users within the cluster. 

For each message $W_i$, we define four binary variables; namely $b_{i,j}, j \in \{i-2,~i-1,~i,~i+1\}$. These are initialized to zero.  
We look at every message starting from $W_1$ to $W_N$ and evaluate the conditions under which a message can be sent and decoded at its desired receiver, such that no interference occurs. If a decision is made to send message $W_{i}$ from transmitter $j$, the corresponding variable $b_{i,j}$ is set to one.
Since message $W_i$ can be sent to its destination using either transmitter $i$ or $i-1$, there are two cases that are considered in the algorithm. In the following we are discussing and justifying both cases. Note that users one and two are considered separately in lines 4-15 since they represent a special case due to their position at the beginning of the cluster.

\textit{Case 1}:  In the first part of the for-loop starting at line 16, we check if message $W_{i}$ can be sent from transmitter $i-1$. This is only possible if message $W_{i}$ is available at transmitter $i-1$ and transmitter $i-1$ does not send $W_{i-1}$. Furthermore, we have to make sure that while sending $W_{i}$, transmitter $i-1$ does not cause interference at receiver $i-1$. There are three possibilities, for which message $W_{i}$ can be decoded without interference. The trivial one is that the link between transmitter $i-1$ and receiver $i-1$ does not exist. Another possible scenario is that receiver $i-1$ is not able to decode its desired message anyway, i.e. $W_{i-1}$ is not sent from transmitter $i-2$. If these conditions are satisfied, then the variable $b_{i,i-1}$ is set to 1. Otherwise, if $W_{i}$ does interfere with $W_{i-1}$ at receiver $i-1$, we might still be able to remove the interference by sending a signal from transmitter $i-2$ such that it will cancel the interference at receiver $i-1$. 

\begin{algorithm}
\caption{}
  \begin{algorithmic}[1]
    \label{alg:the_alg}
\For{i=1:N}
\\ Define $b_{i,i-2}=b_{i,i-1}=b_{i,i}=b_{i,i+1}=0$
\EndFor
\If{$H_{1,1} \neq 0~ \land~ 1 \in \mathcal{T}_{1}$}  
 \State {$b_{1,1}=1$}
\EndIf
\If{$1 \in \mathcal{T}_{2}~\land~ b_{1,1}=0$}  
 \State$b_{2,1}=1$
\ElsIf{$H_{2,2} \neq 0 ~\land~ 2 \in \mathcal{T}_{2}$}  
\If{$b_{1,1}=0$}  
\State$b_{2,2}=1$
\ElsIf {$2\in \mathcal{T}_{1}$}
\State$b_{2,2}=1$; $b_{1,2}=1$
\EndIf
\EndIf
\For{i=3:N}
\If{$(i-1)\in \mathcal{T}_{i}~\land~b_{i-1,i-1}=0$}
\If {$H_{i-1,i-1}=0~\lor~ b_{i-1,i-2}=0$} 
\State{$b_{i,i-1} = 1$}
\ElsIf
 {$(i-2)\in \mathcal{T}_{i}~\land~[H_{i-2,i-2} =0 ~\lor~$ \linebreak $~~~~~~~~~~~~~(b_{i-2,i-2}=0~\land~b_{i-2, i-3}=0)]$}
\State$b_{i,i-1}=1$;  $b_{i,i-2}=1$
\EndIf
\EndIf
~\\
\If {$H_{i,i} \neq 0\land i \in \mathcal{T}_{i}\land b_{i,i-1}=0\land b_{i-2,i-1}=0$}
\If {$b_{i-1,i-1}=0$}
\State$b_{i,i}=1$

\ElsIf {$i \in \mathcal{T}_{i-1}$}
\State$b_{i,i}=1$;
$b_{i-1,i}=1$
\EndIf
\EndIf
\EndFor
\algstore{myalg}
\end{algorithmic}
\end{algorithm}
\begin{algorithm}[H]
	\begin{algorithmic} [1]              
		\algrestore{myalg}
\For{i=1:N}\\
Set $X_{i} = 0$\\
 Generate $X_{i,j},~j \in \{i-1,i\}$ from $W_j$ using an optimal AWGN channel point-to-point code (see e.g.,~\cite{Cover-Thomas})
 \If{$b_{i,i}=1$}
 \State{$X_i \gets X_i + X_{i,i}$}
 \EndIf
 \If{$b_{i+1,i}=1$}
 \State{$X_i \gets X_i + X_{i,i+1}$}
 \EndIf
 \EndFor
 \For{i = 1:N}
 \If{$i \geq 2~\land~b_{i-1,i} = 1$}
 \State{$X_i \gets X_i - \frac{H_{i,i-1} X_{i-1,i-1}}{H_{i,i}}$}
 \EndIf
\If{$i \leq N-2~\land~b_{i+2,i}=1$}
\State{$X_i \gets X_i - \frac{H_{i+1,i+1} X_{i+2,i+1}}{H_{i+1,i}}$}
\EndIf
\EndFor

\end{algorithmic}

\end{algorithm}

This is possible as long as the following conditions hold: First, Message $W_{i}$ must be available at transmitter $i-2$ as well (i.e. $(i-2) \in \mathcal{T}_{i}$). Furthermore, we have to make sure that the signal sent for interference cancellation does not cause interference at receiver $i-2$. This is guaranteed, if either $H_{i-2,i-2} = 0$ or receiver $i-2$ is not able to decode its desired message anyway. In this case, not only $b_{i,i-1}$ but also $b_{i,i-2}$ is set to 1. 


\textit{Case 2}: Now we consider the case of sending message $W_{i}$ from transmitter $i$ (lines 25-31). Here, the trivial conditions to make this possible are that $H_{i,i}$ exists, message $W_{i}$ is available at transmitter $i$, and $W_{i}$ is not being delivered through transmitter $i-1$. This time, we have to make sure that receiver $i$ can decode message $W_{i}$ without any interference. This holds if transmitter $i-1$ is not active. Then $b_{i,i}$ is set to 1. 
Similar to the previous case, we can also cancel the interference from transmitter $i-1$ as long as message $W_{i-1}$ is available at transmitter $i$ and $W_{i-1}$ is the only message that causes interference at receiver $i$. If these conditions hold, both $b_{i,i}$ and $b_{i-1,i}$ are set to 1. 

We now prove the following result to justify Algorithm 1.
\begin{lem}\label{lem:cluster}
	Given any assignment of messages to transmitters, such that each message can only be available at two transmitters, Algorithm 1 leads to the DoF-optimal zero-forcing transmission scheme for users within the input cluster.
\label{lem_m1}
\end{lem}

\begin{IEEEproof}
	
	We consider the messages in ascending order from $W_{1}$ to $W_{N}$, and check which transmitter can deliver message $W_i$ such that it can be decoded at its desired receiver and without interfering at any previous active receiver. If this is true, we will transmit the message. Also, if this is possible through any of the transmitters $i$ and $i-1$, then we prefer to transmit $W_i$ from transmitter $i-1$. In the following, we prove by induction that this procedure leads to the optimal transmission scheme. In a first step, we consider the base case, i.e. we prove that sending $W_1$ from transmitter $1$ is always optimal as long as it is available and the direct link exists. More precisely, as long as $1 \in {\cal T}_1$ and $H_{1,1} \neq 0$.   

We define $\varOmega$ to be the subset of all links $H_{i,j}$ through which a message $W_{i}$, $i\in[K]$ can be sent and decoded at its desired receiver, and call it the feasible set. 
	In other words, all links in $\varOmega$ satisfy the trivial conditions for transmission; namely $j \in {\cal T}_i$ and $H_{i,j} \neq 0$. 
	Let $\mathcal{S} \subset \varOmega\backslash H_{1,1}$ be an arbitrary set of links that can be used \textit{simultaneously} to deliver messages to their desired receivers while eliminating interference. 

Starting with any set ${\cal S}$, if $H_{1,1} \in \varOmega$, we either add $H_{1,1}$ to ${\cal S}$ or replace the first link in ${\cal S}$ by $H_{1,1}$ if there is a conflict. We claim that this replacement cannot decrease the DoF. This is because on one hand, the first active receiver in the network does never observe interference. Also, if we send $W_1$ from the first transmitter, this can only cause interference at the second receiver, but as $H_{2,j},~j \in \{1,2\}$ is either not in $\mathcal{S}$ or it is the first link in $\mathcal{S}$ and hence is replaced by $H_{1,1}$, the transmission of $W_1$ does not prevent any other message corresponding to subsequent links in $\mathcal{S}$ from being decoded at its destination. As a consequence, it is always optimal to transmit $W_1$ from the first transmitter as long as $1 \in {\cal T}_1$ and $H_{1,1} = 1$.
	
Next, we extend the proof to all users by induction. The induction hypothesis is as follows.
We consider an arbitrary link $H_{i,j} \in \varOmega$. Let ${\mathcal{S}}_{1}\subset \varOmega$ be the set of links $H_{k,l}$, through which the subset of messages $\{W_k, k < i\}$ can be delivered simultaneously to their destinations, while eliminating interference. Assume that all links in ${\mathcal{S}}_{1}$ are chosen optimally, i.e. the number of delivered messages cannot be increased by changing any of these links. 


	
Then, we do the induction step. Let ${\mathcal{S}}_{2} \subset \varOmega$ be any set of links $H_{k,l}$, through which a subset of the messages $\{W_{k},k > i\}$ can be transmitted simultaneously such that they can be decoded at their destination. Also, the links in ${\cal S}_2$ are chosen optimally to maximize the number of delivered messages. If it is possible to send $W_{i}$ through $H_{i,i-1}$ without causing a conflict with any of the messages, that are sent through the links in $\mathcal{S}_{1}$, the same logic applies to $H_{i,i-1}$ as to $H_{1,1}$ in the base case. More precisely, if $W_{i}$ does not interfere at any previous active receiver and it can be decoded at receiver $i$ while eliminating interference, $H_{i,i-1}$ can be either added to ${\mathcal{S}}_{2}$ or replace the first link in ${\mathcal{S}}_{2}$, in order to obtain an optimal set of links for the transmission of the messages $\{W_{k},k \geq i\}$. This is possible since again, $W_{i}$ does not cause interference at any active receiver with an index $k > i$, because any of the links $\{H_{i+1,k}, k\in\{i,i+1\}\},$ is either not in ${\mathcal{S}}_{2}$ or it is the link that is replaced by $H_{i,i-1}$. If it is not possible to send $W_i$ through $H_{i,i-1}$ without causing a conflict with any of the messages that are sent through the links in ${\cal S}_1$, but it is possible to do so through $H_{i,i}$, then again the same argument applies for adding $H_{i,i}$ to ${\cal S}_2$. Further, we note that the preference to send $W_i$ through $H_{i,i-1}$ is optimal, since $H_{i,i-1}$ may only cause a conflict with $H_{i+1,i}$ in ${\cal S}_2$, while $H_{i,i}$ may cause a conflict with any of $H_{i+1,i}$ and $H_{i+1,i+1}$.
Therefore, as long as the aforementioned preference rule is applied, sending a message $W_{i}$ through a link $H_{i,j}$ is always optimal as long as it is possible to decode $W_{i}$ at receiver $i$ without causing interference at a previous active receiver. 
	
	This simplifies the optimal algorithm in two ways. On the one hand, we can go through the links one by one and check if it is possible to send a message to its desired receiver without interfering with any of the previous active messages. If it is possible, we will always decide to send the message. On the other hand, decisions that we already made do not have to be changed later, because at each step we make sure to avoid conflicts with previously activated messages. This procedure is applied in Algorithm 1, as we illustrate below.

In the following, we derive the decision conditions for the first three messages in a cluster. \\
If $H_{1,1} \in \varOmega$, sending $W_{1}$ is optimal, as shown in the base case of the proof by induction. Hence, set $b_{1,1} = 1$. 

If $H_{2,1} \in \varOmega$, we have two possibilities. If $b_{1,1} =1$, we cannot send $W_{2}$ from transmitter 1 as well without causing interference at the first receiver. Otherwise, if $b_{1,1} = 0$, it is optimal to send $W_{2}$ from the first transmitter.	

If $H_{2,2} \in \varOmega$ and we are not sending the second message from the first transmitter, i.e., $b_{2,1}=0$, then there are two cases to consider. First, if $b_{1,1}=0$, then $W_1$ is not causing interference at the second receiver and we set $b_{2,2}=1$. Second, if $b_{1,1}=1$, we have interference from $W_1$ at the second receiver. However, this interference can be canceled as long as $W_{1}$ is available at transmitter 2. If this is true, set $b_{2,2}=1$ and $b_{1,2}=1$. 


In the following, we consider sending message $W_{3}$ from the second transmitter if $H_{3,2} \in \varOmega$. We first consider the case where $b_{2,2}=1$. In this case, transmitter $2$ is used to deliver $W_2$ and even if it can be used to deliver $W_3$ as well without causing interference at receiver $2$, this would not increase the sum DoF of the second and third messages, and hence we always set $b_{3,2}=0$ in this case. It hence suffices to only consider the case where $b_{2,2}=0$. There are two cases to consider here. The first is when we can set $b_{3,2}=1$ and no interference cancellation for $W_3$ at the second receiver is needed. This is only possible when either the second receiver is not active, i.e., when $b_{2,1}=0$, or the second direct link is erased, i.e., $H_{2,2}=0$. The second case is when we can set $b_{3,2}=1$ while eliminating the interference caused by $W_3$ at the second receiver by setting $b_{3,1}=1$. This is only possible when $W_3$ is available at transmitter $1$ and the first receiver is not active, i.e., $b_{1,1}=0$.  

Next, we check the possibilities to send $W_{3}$ from the third transmitter if $H_{3,3} \in \varOmega$ and $b_{3,2}=0$. If the second transmitter is inactive, then we set $b_{3,3}=1$. Note that the second transmitter is inactive if $b_{2,2}=0$, since in this case we also know that $b_{1,2}=0$. Otherwise, if $b_{1,2}= 1$, the interference caused by sending $W_1$ cannot be canceled, because $W_1$ is already assigned to the first two transmitters and hence, it cannot be also assigned to transmitter 3. Finally, if $b_{2,2}=1$ and $b_{1,2}=0$, then it is possible to set $b_{3,3}=1$, as long as the interference caused by $W_2$ at the third receiver can be canceled through the third transmitter, which is possible only when $3 \in {\cal T}_2$.


Since each message can only be available at two transmitters, it is not necessary to consider the users before receiver $i-2$ to decide whether $W_i$ can be transmitted. As a consequence, the conditions for sending message $W_3$, if generalized to $W_i$, basically apply to all following messages as well. There is only one additional aspect that have to be considered. If we generalize the case where $b_{i,i-2}=1$ for interference cancellation, we have to make sure that $W_i$ is not causing interference at an active receiver $i-2$. That means for $i-2 > 1$, not only $b_{i-2,i-2} = 0$ but also it is either the case that $b_{i-2, i-3}=0$ or $H_{i-2,i-2}=0$.
\end{IEEEproof}
We now show that Algorithm $1$ can be lead to the optimal zero-forcing scheme in a general $K$-user network. 
\begin{thm}
Algorithm $1$ can be used to achieve the optimal zero-forcing DoF for any realization of a general $K$-user network.

\label{thm_m1}
\end{thm}
\begin{IEEEproof}
If all diagonal links exist, then the whole network is given as input to Algorithm $1$. Otherwise, we scan the diagonal links $H_{i,i-1}, i\in\{2,\cdots,K\}$, in ascending order with respect to the index $i$. Let $i_{\text{min}}$ be the minimum index such that $H_{i,i-1}=0$, then the users with indices $\{1,2,\cdots,i_{\text{min}}-1\}$ form a cluster. For any other index $i$ such that $H_{i,i-1}=0$, let $j$ be the largest index less than $i$ such that $H_{j,j-1}=0$, then the users with indices $\{j,j+1,\cdots,i-1\}$ form a cluster. Finally, for the largest index $i$ such that $H_{i,i-1}=0$, the users with indices $\{i,i+1,\cdots,K\}$ form a cluster. The network is now partioned into clusters; each is given as input to Algorithm $1$ that achieves the optimal zero-forcing DoF within the cluster, which follows from Lemma~\ref{lem:cluster}.

The proof then follows by observing that any assignment of a message outside its cluster can be ignored without loss in optimality. This follows from \cite[Lemma $1$]{ElGamal-Veeravalli-Asilomar13} because no transmitter in a cluster is connected to a receiver outside the cluster. 
\end{IEEEproof}
\section{Simulation}
Using Algorithm 1, we can determine the optimal transmission scheme for a given network realization and message assignment. In this section, we apply this algorithm to compute the DoF as a function of the erasure probability $p$ for several network sizes and message assignments. In particular, we find schemes that outperform those presented in \cite{ElGamal-Veeravalli-Asilomar13} for a wide range of the open interval $0<p<1$.
		
To compute the average puDoF at a certain $p$ for a given message assignment, we simulate a sufficiently large number $n$ of channel realizations, where links are erased with probability $p$, and apply Algorithm 1 to each realization by partitioning the network into clusters as in the proof of Theorem~\ref{thm_m1}. The puDoF value is then computed as the average number of decoded messages divided by the network size $K$. In order to ensure that the computed value holds for large networks, we deactivate the last transmitter in the network, so that if we have a large network that consists of concatenated subnetworks; each of size $K$, then we can achieve the computed puDoF value in the large network by repeating the scheme for each subnetwork, since there will be no inter-subnetwork interference.
		
The simulation is done for a set of message assignments with different fractions $f(p)$ of messages that are assigned to one transmitter connected to their desired receiver and another transmitter that can be used to cancel interference, while the remaining fraction of $1-f(p)$ of messages are assigned to both transmitters that are connected to their destination. Furthermore, we vary the network size $K$. 
More precisely, we use the following assignment strategy:
		\begin{equation*}
		{\scriptsize
			\mathcal{T}_i = 
			\begin{cases}
			\{1, ~2\} & i = 1,\\
			\{K-2,~K-1\} & i = K,\\
			\{i, ~i+1\} & i= 1+ n \cdot \max\left\{2,~\Bigl \lfloor \frac{K}{f(p)\cdot K -1}\Bigr \rfloor\right\},\\
			~& {\tiny n \in \{1,2,\dots, \min\left\{f(p) \cdot K -2,~ \Bigl\lfloor \frac{K}{2} - 1 \Bigr \rfloor\right\}},\\
			\{i, ~i+1\} & i= 2n ,~{\tiny n \in \left\{1,2,\dots,\Bigl\lceil (f(p) - \frac{1}{2} ) K \Bigr\rceil -1\right\}},\\ 
			\{i-1,~ i\} & \text{otherwise,}
			\end{cases}}
			\label{equation:eq1}
		\end{equation*} 	
		where we use the notation $\{1,2, ...,x\}$ to denote the set $[x]$ when $x$ $\geq$1 and the empty set when $x<1$.
	           First, we choose $K$ to be 100 and vary $f(p)$ from $\frac{1}{50}$ up to $\frac{99}{100}$, calculating the puDoF as a function of $p$ for each of these message assignments.
		Additionally, we vary the network size and thus also the message assignment by reducing the fraction by its greatest common divisor. 
		
\begin{figure}[h]	
		\centering
			\includegraphics[width = \columnwidth]{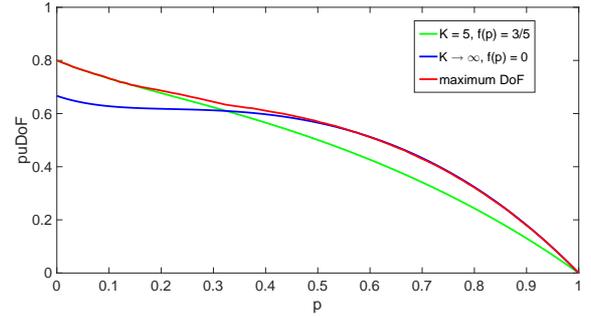}
			\caption{The plot shows the puDoF as a function of the erasure probability $p$ found by applying Algorithm 1 to 6000 randomly generated channel realizations for each value of $p\in\{0,0.01,0.02,\cdots,1\}$. The blue curve corresponds to the message assignment and transmission strategy presented in \cite{ElGamal-Veeravalli-Asilomar13}, which was shown to be optimal as $p \rightarrow 1$. The red curve is the maximum puDoF that is achievable with the message assignments we considered for our simulation.}
			\label{fig:Plot}
		\end{figure}

		As a result, the maximum puDoF that is achievable with the set of message assignments described above is shown in Figure \ref{fig:Plot}. Compared to the schemes presented in \cite{ElGamal-Veeravalli-Asilomar13}, there exist message assignments with a better performance. These are presented in Table \ref{opt_assignments}. Note that in \cite{ElGamal-Veeravalli-Asilomar13} it was shown that assignment with $f(p) = \frac{2}{5}$ is optimal for $p \rightarrow 0$. Interestingly, we find an assignment with $f(p) = \frac{3}{5}$ (see the green curve in Fig. \ref{fig:Plot}) that achieves the same puDoF for $p = 0$, but performs slightly better on the interval $(0, 0.15]$. From our results in Table \ref{opt_assignments}, we observe that the optimal fraction $f(p)$ decreases monotonically from $\frac{3}{5}$ to $0$ as $p$ goes from $0$ to $1$.
		
		\begin{table}[h]
			\begin{centering}
			\begin{tabular}{|c|c|}
				\hline 
				Range of $p$ & Best performing message assignment \\ 
				\hline \hline
				0 to 0.15 & $K = 5,~f(p) = \frac{3}{5}$  \\ 
				\hline 
				0.16 to 0.29 & $K = 100,~ f(p) = \frac{1}{2}$ \\ 
				\hline 
				0.3  & $K = 100,~ f(p) = \frac{49}{100}$ \\ 
				\hline 
				0.31 to 0.32 & $K = 100,~ f(p) = \frac{12}{25}$ \\ 
				\hline 
				0.33 to 0.58 & $K = 100,~ f(p) = \frac{1}{50}$ \\ 
				\hline 
				0.59 to 1 & $K \rightarrow \infty,~ f(p) = 0$ (as in \cite{ElGamal-Veeravalli-Asilomar13}) \\ 
				\hline 
			\end{tabular} 
			\end{centering}
			\caption{Message assignments with the best performance out of the set of assignments that was simulated.}
			\label{opt_assignments}
		\end{table}
		
\bibliographystyle{IEEEtran}

\end{document}